\documentclass[12pt]{article}
\usepackage{amsmath,amssymb,subfigure}
\usepackage[dvips]{epsfig}
\usepackage{amsmath,amssymb,subfigure}
\usepackage[dvips]{epsfig}
\makeatletter \@addtoreset{equation}{section} \makeatother

\begin{document}
\pagestyle{empty}
\newtheorem{theo}{Theorem}[section]
\newtheorem{lem}[theo]{Lemma}
\newtheorem{cor}[theo]{Corollary}
\newtheorem{pro}[theo]{Proposition}
\newtheorem{prop}[theo]{Property}
\newcommand{\no}{\nonumber}
\newcommand{\p}{\partial}
\newcommand{\lb}{\lambda}
\newcommand{\vu}{\vec{u}}
\newcommand{\tvu}{\vec{\tilde{u}}}
\newcommand{\e}{\epsilon}
\newcommand{\ra}{\rightarrow}
\newcommand{\proof}{{\bf Proof}:\quad}

\begin{center}
{\large\bf Large Time Behavior of the Zero Dispersion Limit of the Fifth Order KdV Equation}

\vspace*{.3in}

Virgil Pierce and Fei-Ran Tian

\vspace{.1in}

{\em Department of Mathematics, Ohio State University, Columbus, OH 43210 \\
vpierce@math.ohio-state.edu \ , ~ tian@math.ohio-state.edu}

\vspace*{.1in}

\end{center}

\vspace*{.5in}

\noindent
{\bf Abstract:} \quad
We study the zero dispersion limit of the fifth order KdV equations when time is sufficiently large.
In general, the weak limit may be described by an arbitrary odd number of 
hyperbolic equations. Unlike the KdV case, these are non-strictly hyperbolic equations.
However, 
we show that the weak limit is governed by three hyperbolic equations 
in a domain in the space-time for all times bigger than a large time.
Outside this domain, the weak limit satisfies a single hyperbolic equation.

\thispagestyle{empty}

\newpage

\setcounter{page}{1}
\pagestyle{plain}
%\setcounter{section}{0}
%\section{Introduction}
\refstepcounter{section}
\begin{center}
{\bf $\S$ 1 \quad Introduction}
\end{center}

It is well known that the solution of the KdV equation 
\begin{equation}
u_t + 6 u u_x + \epsilon^2 u_{xxx} = 0  \label{KdV}
\end{equation}
has a weak limit as $\epsilon \rightarrow 0$ while the initial values 
\begin{equation}
u(x, 0; \epsilon) = u_0(x)  \label{ini}
\end{equation}
are fixed \cite{lax, ven}.      
This weak limit satisfies the Burgers equation
\begin{equation}
\label{Burgers}
u_t + (3 u^2)_x = 0
\end{equation}
until its solution develops shocks. Immediately after, the weak limit is governed
by the Whitham equations \cite{fla, lax, ven, whi}
\begin{equation}
\label{KdVW}
u_{it} + \lambda_i(u_1, u_2, u_3) u_{ix} = 0 \ , \quad i=1, 2, 3,
\end{equation}
where the $\lambda_i$'s are given by formulae (\ref{lambda}).
After the breaking of the solution of (\ref{KdVW}), the weak limit is described by systems of
at least five hyperbolic equations similar to (\ref{KdVW}).

The KdV equation (\ref{KdV}) is just the first member of an infinite sequence
of equations, the second of which is the so-called fifth order KdV equation
\begin{equation}
\label{5KdV}
u_t + 30 u^2 u_x + 20 \epsilon^2 u_x u_{xx} + 10 \epsilon^2 u u_{xxx} + \epsilon^4 u_{xxxxx} = 0 \ .
\end{equation}
The solution of the fifth order KdV equation (\ref{5KdV}) also has a weak limit as $\epsilon \to 0$ \cite{lax2}.
As in the KdV case, this weak limit satisfies the Burgers type
equation
\begin{equation}
\label{5Burgers}
u_t + (10 u^3)_x = 0
\end{equation}
until the solution of (\ref{5Burgers}) forms a shock. Later, the limit is governed by equations similar
to (\ref{KdVW}), namely,
\begin{equation}
\label{5KdVW}
u_{it} + \mu_i(u_1, u_2, u_3) u_{ix} = 0 \ , \quad i=1, 2, 3,
\end{equation}
where $\mu_i$'s are given in (\ref{mu}). They will be also be called the Whitham equations \cite{pie}.

In this paper, we are interested in the large time behavior of the weak limit of
the fifth order KdV equation (\ref{5KdV}). 

For simplicity,
we consider $u_0(x)$ of (\ref{ini}) which is
a decreasing function and is bounded at $x= \pm \infty$:
\begin{equation}
\label{cu}
\lim_{x \rightarrow - \infty }u_0(x) = 1 \ , ~~~~
\lim_{x \rightarrow + \infty }u_0(x) = 0 \ .
\end{equation}
   
The large time behavior of the weak limit of the KdV equation (\ref{KdV}) 
has been studied \cite{Tian3}.
For generic initial data of (\ref{cu}), the weak limit is governed by
the Whitham equations (\ref{KdVW}) in a domain in the space-time when $t$ is sufficiently
large. Outside this domain, the weak limit satisfies the Burgers equation (\ref{Burgers}).
We note that the weak limit may be described by an arbitrary odd number of hyperbolic equations
in the intermedia times.

The strong hyperbolicity of the Whitham equations (\ref{KdVW}) for the KdV 
plays an important role in the paper of \cite{Tian3}.
It is well known that equations (\ref{KdVW}) are strictly hyperbolic:
$$ \lambda_i(u_1, u_2, u_3) \neq  \lambda_j(u_1, u_2, u_3) \ , \quad i, j = 1, 2, 3; i \neq j,$$ 
and genuinely nonlinear:
$$ {\p \over \p u_i} \lambda_i(u_1, u_2, u_3) \neq 0 \ , \quad i = 1, 2, 3, $$
for $u_1 > u_2 > u_3$ \cite{lev}.

However, in the case of fifth order KdV equation (\ref{5KdV}), the Whitham equations (\ref{5KdVW})
are neither strictly hyperbolic nor genuinely nonlinear \cite{pie}. 

In this paper, we use the method developed in \cite{gra1} and \cite{gra2} to study the weak limit of
the fifth order KdV (\ref{5KdV}) for sufficiently large time $t$. Our approach does not require 
the Whitham equations to be strongly hyperbolic. Our result is quite similar to the 
KdV result. Namely, For generic initial data of (\ref{cu}), the weak limit is governed by
the Whitham equations (\ref{5KdVW}) in a domain $x^-(t) < x < x^+(t)$ in the space time 
when $t$ is sufficiently
large (see Figure 1.). Outside this domain, the weak limit satisfies the Burgers type
equation (\ref{5Burgers}).

We also find some difference between the KdV case and the fifth order KdV case. In the latter case,
the trailing edge is ``lifted'', i.e., 
$$u_1 \approx 1 \ , \quad u_2 = u_3 \approx {1 \over 4}  \quad \mbox{for large $t>0$}$$
at the trailing edge $x=x^-(t)$ (see Figure 1.). In the KdV case, instead we have
$$u_1 \approx 1 \ , \quad u_2 = u_3 \approx 0  \quad \mbox{for large $t>0$}$$
at the trailing edge \cite{gur, lax}.

This phenomenon of ``lifted'' trailing edge has also been observed in the weak limit of 
the fifth order KdV equation
when the initial values $u_0(x)$ of (\ref{ini}) are given by a step-like function
$$u_0(x) = \left\{ \begin{matrix} 1 & x < 0 \\
0 & x > 0 \end{matrix} \right. \ . $$
In this case, we have  
$$u_1 = 1 \ , \quad u_2 = u_3 = {1 \over 4}$$
at the trailing edge $x=-15t$ \cite{pie}.

The organization of this paper is as follows. In Section 2, we will summarize the method
of \cite{gra1}. In Section 3, we will describe our main theorem. In Section 4, we
will study the trailing and leading edges, which separate the Burgers type solutions and the
Whitham solutions,
when $t$ is large. In Section 5, we will study the 
Whitham solutions that live between the trailing and leading edges.

\refstepcounter{section}
\begin{center}
{\bf $\S$ 2 \quad A Minimization Problem}
\end{center}

As in the KdV case, the weak limit of the fifth order KdV equation (\ref{5KdV})
is determined by
a minimization problem with constraints \cite{lax2}.

The minimization problem is
\begin{equation}
\label{mini}
\underset{\{\psi \geq 0, \  \psi \in L^1 \}}
{\rm
Minimize} 
\{ - \frac{1}{2 \pi} \int_0^1 \int_0^1 \log \Big|\frac{\eta - \mu}
{\eta + \mu
}\Big|
\psi(\eta) \psi(\mu) d \eta d \mu + \int_0^1 a(\eta, x, t)
\psi(\eta) d \eta \} \ .
\end{equation}
Function $a(\eta, x, t)$ is given by
\begin{eqnarray}
a(\eta, x, t) &=& \eta x - 16 \eta^5 t - \theta(\eta) \ , \label{a} \\
\theta(\eta) &=& \eta f(\eta^2) + \int_{f(\eta^2)}^{+ \infty}
[\eta - \sqrt{\eta^2 - u_0(x)}] d x \ , \no
\end{eqnarray}
where $f(u)$ is the inverse function of the initial data $u_0(x)$.
We note that in the KdV case, we have $a(\eta, x, t) = \eta x - 4 \eta^3 t - \theta(\eta)$
instead of (\ref{a}) \cite{gra1, lax, ven}. 

We now give a brief summary of our approach to the minimization problem (\ref{mini}).

We first follow Lax $\&$ Levermore \cite{lax} and Venakides \cite{ven, ven2} to make the ansatz
that the support of $\psi$ consists
of a finite union of disjoint intervals
\begin{equation}
\label{support}
S = [0, \sqrt{u_{2g+1}}] \cup [\sqrt{u_{2g}}, \sqrt{u_{2g-1}}] \cup \cdots
\cup [\sqrt{u_2}, \sqrt{u_1}]  \ ,
\end{equation}
where $0 < u_{2g+1} < \cdots < u_2 < u_1 < 1$. Hence, $g$ is the number of gaps in the support
$S$.
 
We then introduce 
\begin{equation}
\label{P}
P(\xi,\vec{u}) = 2 R^2(\xi, \vu) \Phi_g(\xi,\vec{u}) + Q(\xi,\vec{u}) \ ,
\end{equation}
where $\vu$ denotes $(u_1, u_2, \cdots, u_{2g+1})$ and
$R(\xi, \vec{u}) = \sqrt{(\xi - u_1)\cdots (\xi - u_{2g+1})}$ with the sign
given by $\sqrt{1} = 1$ \cite{gra1}. 

The function $\Phi_g(\xi,\vec{u})$ 
is the unique solution of the boundary value problem for
the Euler-Poisson-Darboux equations
\begin{eqnarray}
2(u_i - u_j) {\p^2 \Phi_g \over \p u_i \p u_j} &=& {\p \Phi_g \over \p u_i} -
{\p \Phi_g \over \p u_j} \ , \quad i,j = 1, 2, \cdots, 2g+1, \label{ph1} \\
2(\xi - u_i) {\p^2 \Phi_g \over \p \xi \p u_i} &=& {\p \Phi_g \over \p \xi} -
2{\p \Phi_g \over \p u_i} \ , \quad i = 1, 2, \cdots, 2g+1, \label{ph2} \\
\Phi_g(u, u, \cdots, u) &=& {2^g \over (2g+1)!!} {d^{g+1} \over d u^{g+1}}
[ x - 30 t u^2 - f(u)] 
\ . \label{ph3}
\end{eqnarray}

The function $Q(\xi,\vec{u})$ is a polynomial of degree $2g$ in $\xi$: 

\begin{enumerate}

\item for $g=0$,
$Q(\xi, u_1) = x - 30 u_1^2 t - f(u_1)$;

\item for $g > 0$,
\begin{eqnarray*}
Q(\xi, \vec{u}) &=& 2 \sum_{i = 1}^{2g+1}
[\prod_{l=1, l \neq i}^{2g+1} (\xi-u_l)]
{\p q_{g,g}(\vec{u}) \over \p u_i} \no \\
& & + \sum_{k=1}^{g} [(2k-1) \sum_{l=0}^{g-k} \Gamma_l
(\vec{u})q_{g,k+l}(\vec{u})]
P_{g, k-1}(\xi, \vec{u}) \ . 
\end{eqnarray*}
\end{enumerate}
Here $\Gamma_{l}(\vec{u})$ are the coefficients of the
expansion at $\xi = \infty$
\begin{displaymath}
R(\xi, \vu) = \xi^{g+{1 \over 2}} [\Gamma_{0}(\vu) +
{\Gamma_{1}(\vu) \over
\xi}  + {\Gamma_{2}(\vu) \over \xi^2} + \cdots ]
\end{displaymath}
and $q_{g,k}(\vec{u})$ is the solution of the boundary value problem for
another version of the Euler-Poisson-Darboux equations
\begin{eqnarray}
2(u_i - u_j) {\p^2 q_{g,k} \over \p u_i \p u_j} = {\p q_{g,k} \over \p u_i} -
{\p q_{g,k} \over \p u_j} \ , \quad \mbox{$i, j= 1, \cdots, 2g+1$} \ , \label{q1} \\
q_{g,k}(u, \cdots, u) = \frac{ 2^{g-1}}{(2g-1)!!}
u^{-k + \frac{1}
{2}} \frac{d^{g-k} }{d u^{g-k}} \{ u^{g - \frac{1}{2}} [ \frac{
d^{k-1} }{d u^{k-1}} \left( x - 30 t u^2 - f(u) \right) ] \}. \label{q2}
\end{eqnarray}
The polynomial 
\begin{equation}
\label{pgn}
P_{g,n}(\xi,\vec{u}) = \xi^{g+n} + a_{g,1} \xi^{g+n-1} + \cdots + a_{g, g+n}
\end{equation}
is defined such that
\begin{equation}
\label{Pgn}
\frac{P_{g,n}(\xi,\vec{u})}{R(\xi, \vu)} =
\xi^{n-\frac{1}{2}} +
O(\xi^{-\frac{3}{2}}) ~~~~~~~
\mbox{for large $|\xi|$} \ ,
\end{equation}
and
\begin{equation}
\label{lc}
\int_{u_{2k+1}}^{u_{2k}} \frac{P_{g,n}(\xi,\vec{u})}{R(\xi, \vu)} d \xi
= 0 \ , ~~~~~~~~
k = 1, 2, \cdots, g \ ,
\end{equation}
where $\vu=(u_1, u_2, \cdots, u_{2g+1})$.

Solutions of the boundary value problem for the Euler-Poisson-Darboux equations
can be constructed using those of the following simpler problem as building
blocks \cite{Tian1}
\begin{eqnarray}
\label{EPD}
2(x_{1} - x_{2})\frac{\partial^{2} v}{\partial x_{1} \partial
x_{2}} & = & \frac{\partial v}{\partial x_{1}} - \rho \frac{\partial v}
{\partial x_{2}} \ , ~~~~~ \rho > 0  ~ is ~ a ~ constant \ , \\
v(x_1, x_1) & = & g(x_1) \ . \label{EPDd}
\end{eqnarray}
The boundary value problem (\ref{EPD}-\ref{EPDd}) has one and
only one solution for smooth boundary
data $g(x)$ \cite{Tian1}. A simple calculation shows that the solution is
given by the formula
\begin{displaymath}
v(x_1, x_2) = C \int_{-1}^{1} \frac{g(\frac{1+\mu}{2}x_1 + \frac{1-\mu}{2}x_2)}{
\sqrt{1 - \mu^{2}}} (1 + \mu)^{\frac{\rho - 1}{2}} d \mu \ ,
\end{displaymath}
where
\begin{displaymath}
C = \frac{1}{\int_{-1}^{1} \frac{(1+\mu)^{\frac{\rho - 1}{2}}}{\sqrt{1-\mu^{2}}}
d \mu} \ .
\end{displaymath}
A change of integration variable gives another
expression for the solution
\begin{equation}
\label{EPQsS}
v(x_1,x_2) =   C \left[{2 \over x_1-x_2}\right]^{\rho - 1 \over 2} \int_{x_2}
^{x_1} g(x) {(x - x_2)^{\rho - 2 \over 2} \over (x_1 - x)^{1 \over 2}} d x \ ,
\end{equation}
where the square root is set to be positive for $x$ between $x_1$ and $x_2$.
In particular, when $g=0$, the solution of the boundary value problem
(\ref{ph1}-\ref{ph3}) is
\begin{equation}
\label{phi0}
\Phi_0(\xi, u_1) = {1 \over 2 \sqrt{2}} \int_{-1}^{1} {H'({1 + \mu \over 2}
\xi + {1- \mu \over 2} u_1) \over \sqrt{1 - \mu} } d \mu \ ,
\end{equation}
where $H(u) = x - 30 t u^2 - f(u)$.
More generally, the solution of equations (\ref{ph1}-\ref{ph3}) can be solved explicitly
\begin{eqnarray}
\lefteqn{\Phi_g(\xi, \vec{u}) = } \no \\
&M& \int_{-1}^{1} \cdots \int_{-1}^{1}
{ H^{(g+1)}({1 + \mu_{2g+1} \over 2} (\cdots ({1 + \mu_{2} \over 2}
({1 + \mu_{1} \over 2} \xi + {1 - \mu_{1} \over 2}u_1) + \cdots ) +
{1 - \mu_{2g+1} \over 2} u_{2g+1} ) \over \sqrt{(1-\mu_1)(1-\mu_2) \cdots
(1 - \mu_{2g+1})}} \no \\
& & \times (1+\mu_2)^{1 \over 2} (1+\mu_3)^{2 \over 2} \cdots
(1+\mu_{2g+1})^{2g \over 2} d \mu_1 d \mu_2 \cdots d \mu_{2g+1} \ , \label{phi}
\end{eqnarray}
where the constant $M$ is chosen so that 
the boundary condition (\ref{ph3}) is
satisfied \cite{gra1}. 

The solution $q_{g,k}(\vec{u})$ of another
boundary value problem
(\ref{q1}), (\ref{q2}) is also given by a multiple integral formula similar to (\ref{phi}).

We now list some identities of $\Phi_g$ and $q_{g,k}$. They will be useful
in the subsequent calculations. 

\begin{lem} \cite{gra1}
\label{gidentity}
\begin{eqnarray}
2 (\xi - u_i) {\p \over \p u_i} \Phi_g(\xi, \vec{u})
&=& \Phi_g(\xi, \vec{u}) - \Phi_g(u_i, \vec{u}) \ , \label{Phiu} \\
\Phi_g(\xi, \vu)|_{u_l=u_{l+1} = u^*} &=&
\frac{\Phi_{g-1}(\xi,\tvu)-\Phi_{g-1}(u^*,\tvu)}{\xi-u^*} \ , \quad 
\label{lim1} \\
\Phi_g(u^*,\vec{u})|_{u_l=u_{l+1} = u^*} &=& {\p \Phi_{g-1}(\xi, \tvu)
\over \p \xi}|_{\xi
= u^*} \ , \label{lim2} \\
\Phi_{g-1}(\xi,\tvu) &=& \left(\sum_{i=1}^{2g+1}
{\p \over \p u_i}q_{g,g}(\vu)\right)
|_{u_{l}=u_{l+1}=\xi} \ ,  \label{lim3}
\end{eqnarray}
where $\vu=(u_1,\dots, ,u_{2g+1})$,
$\tvu=(u_1,\dots, u_{l-1},u_{l+2},\dots,u_{2g+1})$ and $1\leq l\leq 2g$.
\end{lem}

Our method of solution to the minimization problem is summarized in the 
following
theorem.
\begin{theo} \cite{gra1}
\label{main}
If $x$, $t$ and $\vu$ are connected by the equations
\begin{equation}
\label{P2}
P(u_i, \vu) = 0 ~~~~~~ \mbox{for $i=1, 2, \cdots, 2g+1$},
\end{equation}
where $P$ is defined by (\ref{P}), and if the inequalities for 
$k=0, 1, \cdots, g$, $u_0 = 1$ and $u_{2g+2} = 0$
\begin{eqnarray}
Re\{\sqrt{-1}R(\xi,\vu)\}\Phi_g(\xi,
\vu) & < & 0 \quad
\mbox{for $u_{2k+2} < \xi < u_{2k+1}$} \ , \label{constraint} \\
\int_{u_{2k+1}}^{\xi} R(\mu, \vu) \Phi_g(\mu,\vu)
d \mu  & > &  0 \quad \mbox{for $u_{2k+1}<\xi<u_{2k}$} \ ,
\label{constraint2}
\end{eqnarray}
are satisfied, the function 
\begin{equation}
\label{psi}
\psi(\eta) = - 2 \eta Re\{\sqrt{-1}R(\eta^2,\vu)\}\Phi_g(\eta^2,\vu) 
\end{equation}
is the minimizer of (\ref{mini}). Its support is given by
(\ref{support}).
\end{theo}

We now analyze equations (\ref{P2}). The boundary conditions 
(\ref{ph3}) and (\ref{q2}) are linear in $x$ and $t$, as is the function
$P(\xi, \vec{u})$. We then use formulae (\ref{P}-\ref{q2}) to write 
\begin{equation}
\label{1}
P(\xi, \vec{u}) = x P_{g,0}(\xi, \vec{u}) - 80 t  P_{g,2}(\xi, \vec{u}) - P_f(\xi, \vec{u}) \ ,
\end{equation}
where $P_{g,0}$ and $P_{g,2}$ are defined by (\ref{pgn}-\ref{lc}), and
$P_f$ is also determined by equations (\ref{P}-\ref{q2}) with boundary data (\ref{ph3})
and (\ref{q2}) depending only on $f$ and its derivatives. 

Equations (\ref{P2}) can then be written as
\begin{equation}
\label{P3}
x = \lambda_{g, i}(\vec{u})t + w_{g,i}(\vec{u})  \ , \quad i = 1, 2, \cdots , 2g+1 \ ,
\end{equation}
where $$  \lambda_{g, i} = 80 {P_{g,2}(u_i, \vec{u}) \over P_{g,0}(u_i, \vec{u})} \ ,
\quad w_{g,i}(\vec{u}) = {P_f(u_i, \vec{u}) \over P_{g,0}(u_i, \vec{u})} \ .$$

It is well known that the solutions $u_1$, $u_2$, $\cdots$, $u_{2g+1}$ of (\ref{P3})
or equivalently (\ref{P2}), as functions of $x$ and $t$, satisfy an odd number of 
hyperbolic partial differential 
equations \cite{dub, gra1, kri, Tian2, tsa}
\begin{equation}
\label{P4}
u_{it} + \lambda_{g, i}(u_1, u_2, \cdots, u_{2g+1}) u_{ix} = 0 \ , \quad i = 1,2, \cdots, 2g+1 \ .
\end{equation}
In other words, the square roots of the end points of the support (\ref{support}) of the minimizer of 
the variational problem (\ref{mini}) satisfy the PDE (\ref{P4}).

The coefficients $\lambda_{g,i}$ of (\ref{P4}) involve complete hyperelliptic integrals
of genus $g$ because $P_{g,n}$'s do so in view of (\ref{pgn}-\ref{lc}). Equations (\ref{P4}) are then 
called the $g$-phase Whitham equations and its solution a $g$-phase solution.
 
In particular, when $g=0$, we have $\lambda_{0, 1}(u_1) = 30 u_1^2$ and equation (\ref{P4}) becomes
the Burgers type equation (\ref{5Burgers}). 
   
When $g=1$, equations (\ref{P4}) turn out to be the Whitham equations (\ref{5KdVW}), where
\begin{equation}
\label{mu}
\mu_i(u_1, u_2, u_3) = \lambda_{1, i}(u_1,u_2,u_3) = 80 {P_{1,2}(u_i, u_1, u_2, u_3) 
\over P_{1,0}(u_i, u_1, u_2, u_3)} \ , \quad i=1,2,3 \ .
\end{equation}

We note that, for the KdV equation, $\lambda_i$'s of (\ref{KdVW}) are given by
\begin{equation}
\label{lambda}
\lambda_i(u_1, u_2, u_3) = 12 {P_{1,1}(u_i, u_1, u_2, u_3) \over P_{1,0}(u_i, u_1, u_2, u_3)} \ ,
\quad i = 1, 2, 3 \ . 
\end{equation}

To solve the system of algebraic equations (\ref{P2}), we rely on the
Implicit Function Theorem. Although the system is complicated, its Jacobian
matrix is diagonal and its determinant is easily calculated.
\begin{theo} \cite{gra1}
\label{jaco}
On the solution $\vec{u}$ of (\ref{P2}), 
\begin{enumerate}
\item the Jacobian matrix of the hodograph transform (\ref{P2}) is diagonal, i.e.,
$${\p P(u_i, \vec{u}) \over \p u_j} = 0 \ , \quad i, j = 1, 2, \cdots, 2g+1 \ ; i \neq j \ ,$$

\item the Jacobian is 
\begin{displaymath}
det \left({\p P(u_i, \vec{u}) \over \p u_j} \right)
= (-1)^g  \prod_{i > j} (u_i - u_j)^2 \prod_{i=1}^{2g +1}
\Phi_g(u_i, \vu)  \ .
\end{displaymath}
\end{enumerate}
\end{theo}

Violation of inequalities (\ref{constraint}-\ref{constraint2}) gives rise to
phase changes. Indeed,
when (\ref{constraint}) first fails, function $\Phi_g(\xi,\vu)$
must have a double $\xi$-zero for $\sqrt{\xi}$ in the interior of $S$ of (\ref{support}),
say, the zero $u^*$ where $\sqrt{u^*} \in S$.
The resulting equations are
\begin{eqnarray}
\Phi_g(u^*, \vu)
&=& 0 \ , \label{teg1} \\
{\p \over \p \xi} \Phi_g(\xi, \vu)|_{
\xi = u^*} &=& 0  \ , \label{teg2} \\
P(u_i, \vu) &=& 0 \ , \quad i = 1, 2, \cdots, 2g+1 \ . 
\label{teg3}
\end{eqnarray}
They govern the 
so-called trailing edge separating
$g$ and $g+1$ phases.
Similarly, when (\ref{constraint2}) is first violated, the
indefinite integral must also
have a double $\xi$-zero for $\sqrt{\xi}$ in the complement of $S$,
say, the zero $u^* \in (u_{2k+1}, u_{2k})$.
The resulting equations are
\begin{eqnarray}
\Phi_g(u^*, \tvu)
&=& 0 \ , \label{leg1} \\
\int_{u_{2k+1}}^{u^*} \Phi_g(\xi, \vu)
R(\xi, \vu) d \xi &=& 0 \ , \label{leg2} \\
P(u_i, \vu) &=& 0 \ , \quad i = 1, 2, \cdots, 2g+1 \ ,
\label{leg3}
\end{eqnarray}
which govern
the so-called leading edge separating $g$ and $g+1$ phases.

We close this section with a remark on the other higher order KdV equations. 
All the zero dispersion limits of the equations in the KdV hierarchy are determined by
the minimization problem (\ref{mini}) with $a(\eta, x, t)$ of (\ref{a}) replaced by \cite{lax2}
\begin{equation}
\label{am}
a(\eta, x, t) = \eta x - 4^m \eta^{2m+1} t - \theta(\eta) 
\end{equation}
for $m=1 \ , 2 \ , \cdots \ .$ The function $x-30tu^2 - f(u)$ in the 
boundary conditions (\ref{ph3}) and (\ref{q2}) 
must then be replaced by $x - C_m t u^m - f(u)$, where \cite{gra1}  
$$C_m = {(2m+1)2^{2m+1} \over \pi} \int_0^1 {t^{2m} \over \sqrt{1 - t^2}} d t \ .$$
In particular, $m=1$ corresponds to the KdV case and $m=2$ to the fifth order KdV case.
All the analysis of Section 3 can be carried over from the fifth order KdV ($m=2$) case to all 
the other equations ($m \neq 2$) in the KdV hierarchy.  

In the subsequent sections, we will only study the fifth order KdV ($m=2$) case. The KdV
($m=1$) case has been well understood \cite{Tian3}. We will indicate why the 
$m > 2$ case is technically more difficult than the $m=2$ case when the opportunity presents itself.

\bigskip

\refstepcounter{section}
\begin{center}
{\bf $\S$ 3 \quad Main Theorem}
\end{center}

We first make assumption on the initial data $u_0(x)$ or its inverse
$f(u)$. Since $u_0(x)$ has the limits (\ref{cu}) as $x$ goes to the infinity,
its inverse $f(u)$ behaves as
\begin{equation}
\label{cf1}
\lim_{u \rightarrow 0} f(u) = + \infty \ , ~~~~~~~~~~~~~~~~~
\lim_{u \rightarrow 1} f(u) = - \infty \ .
\end{equation}
We further assume that 
\begin{equation}
\label{cf2}
f'''(u) < 0 ~~~~~~~~ \mbox{in the neighborhood of $u=0$ and $u=1$} \ .
\end{equation}
Conditions (\ref{cf1}-\ref{cf2}) immediately imply \cite{Tian3}
\begin{equation}
\label{cf3}
\lim_{u \rightarrow 0} f'(u) = \lim_{u \rightarrow 1} f'(u) = - \infty \ , ~~
\lim_{u \rightarrow 0} f''(u) = + \infty \ , ~ 
\lim_{u \rightarrow 1} f''(u) = - \infty \ .
\end{equation}

\begin{theo} (see Figure 1.)
\label{Main}
Under the conditions (\ref{cf1}) and (\ref{cf2}), there exists a $T>0$ so that when 
$t > T$, the weak 
limit of the fifth order KdV equation is governed by the Whitham equations
(\ref{5KdVW}) for $x^-(t) < x < x^+(t)$, where $x^-(t)$ and $x^+(t)$ are some functions of $t$.
The weak limit satisfies the Burgers
type equation (\ref{5Burgers}) for $x < x^-(t)$ and $x > x^+(t)$.
\end{theo}

This theorem is a consequence of Theorems \ref{Burgers2} and \ref{Whitham}, which will be
proved in the subsequent sections.  

\begin{figure}[h] \label{fig1}
\begin{center}
\resizebox{15cm}{5cm}{\includegraphics{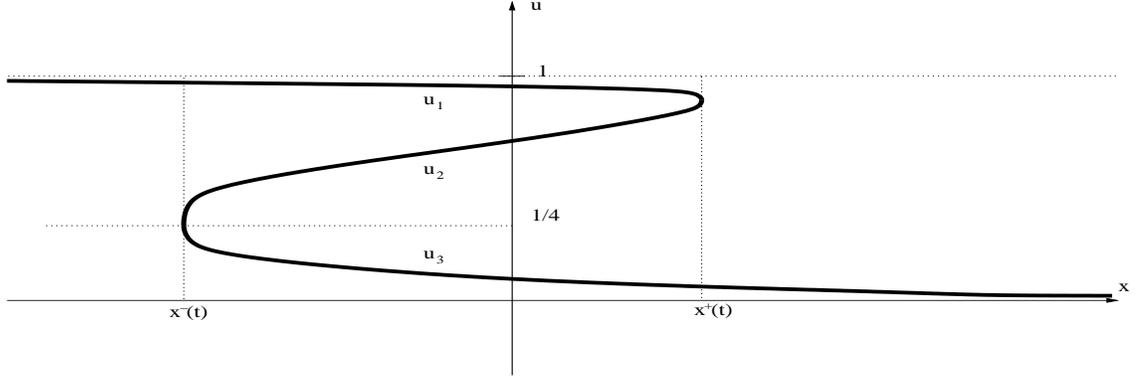}}
\caption{Large time behavior of the weak limit of the fifth order KdV
for generic initial data.
The limit is governed by
the Whitham equations (\ref{5KdVW}) for $x^-(t) < x < x^+(t)$ and by
the Burgers type equation (\ref{5Burgers}) otherwise. The trailing edge $x=x^-(t)$ is
characterized by $u_1 \approx 1 \ , u_2=u_3 \approx 1/4$.}
\end{center}\end{figure}

\bigskip

\refstepcounter{section}
\begin{center}
{\bf $\S$ 4 \quad Trailing and Leading Edges}
\end{center}

In this section, we will study the trailing and leading edges of a single phase
solution when $t$ is sufficiently large.

\medskip

\noindent
{\bf $\S$ 4.1 \quad The trailing edge}

We first analyze the tail of a single phase solution at which $u_2=u_3$ (see Figure 1.).
It is a phase transition boundary between zero and single phases; it corresponds to
$g=0$ in equations (\ref{teg1}-\ref{teg3}):
\begin{eqnarray}
W_1(u_1, u_3, t) := \Phi_0(u_3, u_1)
&=& 0 \ , \label{teg10} \\
W_2(u_1, u_3, t) := {\p \over \p \xi} \Phi_0(\xi, u_1)|_{
\xi = u_3} &=& 0  \ , \label{teg20} \\
x - 30 t u_1^2 - f(u_1) &=& 0 \ . \label{teg30}
\end{eqnarray}

We split $\Phi_0(\xi, u_1)$ into
\begin{equation}
\label{split}
\Phi_0(\xi, u_1) = - 30 t U_0(\xi, u_1) - F_0(\xi, u_1) \ , 
\end{equation}
where $U_0$ and $F_0$ are the solutions of the Euler-Poisson-Darboux equations
(\ref{ph1}-\ref{ph3}) with boundary data $2 u$ and $f'(u)$, respectively.
In view of formula (\ref{phi0}), we have
\begin{eqnarray}
F_0(\xi, u_1) &=& {1 \over 2 \sqrt{2}} \int_{-1}^{1} {f'({1 + \mu \over 2}
\xi + {1- \mu \over 2} u_1) \over \sqrt{1 - \mu} } d \mu  \ , \label{fF} \\
U_0(\xi, u_1) &=& {1 \over 2 \sqrt{2}} \int_{-1}^{1} { 2({1 + \mu \over 2}
\xi + {1- \mu \over 2} u_1) \over \sqrt{1 - \mu} } d \mu = 
{4 \over 3} \xi + {2 \over 3} u_1 \ . \label{fU}
\end{eqnarray}

Equations (\ref{teg10}) and (\ref{teg20}) are then equivalent to
\begin{eqnarray}
{F_0 \over U_0} + 30 t &=& 0
\ , \label{teg10'} \\
{\p \over \p \xi} [{F_0 \over U_0} + 30 t]_{\xi = u_3}
&=& 0  \ . \label{teg20'}
\end{eqnarray}

Equation (\ref{teg20'}) suggests that the point $\xi = u_3$ is a critical
point of the function $[F_0/U_0 + 30 t]$.
We therefore consider an auxiliary minimum problem
\begin{equation}
\label{min}
Min_{0 < \xi < 1} \left[ - {F_0(\xi, u_1) \over U_0(\xi, u_1)} \right] 
\end{equation}
for each $u_1$ close to $1$.    

To study the minimization problem (\ref{min}), we need the following lemma.
\begin{lem}
\label{trailing}

\begin{enumerate}

\item 
\begin{equation}
\label{F}
\lim_{\xi \rightarrow 0} F_0(\xi, u_1) = \lim_{\xi \rightarrow 1} F_0(\xi, u_1)
= - \infty 
\end{equation}
for each $u_1 \in (0, 1)$.

\item There exists a $\delta > 0$ such that 
\begin{equation}
\label{F_0''}
{\p^2 \over \p \xi^2} F_0(\xi, u_1) < 0 ~~~~~ \mbox{for $0 < \xi < 1$ and 
$1 - \delta < u_1 < 1$} \ .
\end{equation}

\item 
\begin{equation}
\label{F'} 
\lim_{u_1 \rightarrow 1} { {\p F_0(\xi, u_1) \over \p \xi} \over F_0(\xi, u_1) } = {1 \over 1 - \xi}
\end{equation}
uniformly for $\xi$ on every compact subset of the interval $(0,1)$.

\end{enumerate}
\end{lem}
Proof: To prove the first limit of (\ref{F}), we use (\ref{fF}) to 
rewrite $F_0(\xi, u_1)$ as
$$F_0(\xi, u_1) = {1 \over 2 \sqrt{\xi - u_1}} \int_{u_1}^{\xi} {f'(u)
\over \sqrt{\xi - u} } d u ~~~~~ \mbox{when $\xi > u_1$} \ . $$
Since $f'(u) < 0$, we estimate the integral when $\xi$ is close to $1$:
\begin{equation*}
F_0(\xi, u_1) \leq {1 \over 2(\xi - u_1)} \int_{u_1}^{\xi} f'(u) d u
= {1 \over 2(\xi - u_1) } [ f(\xi) - f(u_1) ] \ . 
\end{equation*}
This together with (\ref{cf1}) proves the second limit of (\ref{F}).
The first limit can be shown in the same way.

To prove inequality (\ref{F_0''}), in view of condition (\ref{cf2}) we
assume $f'''(u) < 0$ for $u$ outside the interval $(\delta_1, 1 - \delta_1)$
where $\delta_1$ is a small positive number.
Since $f'''(u) < 0$ for $u > 1 - \delta_1$, it follows from formula
(\ref{fF}) that it suffices to prove (\ref{F_0''}) for $\xi \leq 1 - \delta_1$.

We use formula (\ref{fF}) to write for $u_1 > 1 - {\delta_1 \over 2} \geq \xi$,
\begin{eqnarray}
{\p^2 \over \p \xi^2} F_0(\xi, u_1) &=& {1 \over 2 (u_1 - \xi)^{5 \over 2}}
\int_{\xi}^{u_1} f'''(u) {(u_1 - u)^2 \over \sqrt{u - \xi} } d u \no \\
&=& {1 \over 2 (u_1 - \xi)^{5 \over 2}} [ \int_{1 - {\delta_1 \over 2}}^{u_1}
+ \int_{\xi}^{1 - {\delta_1 \over 2}} ] \ . \label{F0''}
\end{eqnarray}
Since $f'''(u) < 0$ for $u < \delta_1/2$, 
the second integral in the parenthesis of
(\ref{F0''}) is smaller than
$$\int_{\delta_1}^{1 - {\delta_1 \over 2}} {|f'''(u)| \over \sqrt{u - \xi}}
d u \ ,$$
which, since $f'''(u)$ is bounded on the closed interval $[\delta_1, 1 - \delta_1/2]$, 
is uniformly bounded for $\xi \leq 1 - \delta_1/2$.

Since $f'''(u) < 0$ for $u > 1 - \delta_1$, the first integral of (\ref{F0''})
is less than
\begin{eqnarray*}
\lefteqn{\int_{1 - {\delta_1 \over 2}}^{u_1}
f'''(u) (u_1 - u)^2 d u} \\
&=& 2 f(u_1) -
f''(1 - {\delta_1 \over 2}) (u_1 - 1 + {\delta_1 \over 2})^2 -
2 f'(1 - {\delta_1 \over 2}) (u_1 - 1 + {\delta_1 \over 2}) -
2f(1 - {\delta_1 \over 2}) \ ,
\end{eqnarray*}
which, in view of (\ref{cf1}), goes to $- \infty$ as $u_1 \rightarrow
1$. This proves inequality (\ref{F_0''}).

To prove (\ref{F'}), we
use formula (\ref{fF}) to write for $u_1 > \xi$,
\begin{eqnarray}
F_0(\xi, u_1) &=& {1 \over 2 \sqrt{u_1 - \xi}} \int_{\xi}^{u_1} {f'(u)
\over \sqrt{u - \xi } } d u \ ,  \label{fF2} \\
{\p \over \p \xi} F_0(\xi, u_1) &=& {1 \over 2 (u_1 - \xi)^{3 \over 2}}
\int_{\xi}^{u_1} f''(u) {u_1 - u \over \sqrt{u - \xi} } d u  \label{fF2'} \ .
\end{eqnarray}

We first observe that 
\begin{equation}
\label{f'}
\lim_{u_1 \rightarrow 1} \int_{\xi}^{u_1} {f'(u)
\over \sqrt{u - \xi } } d u = - \infty
\end{equation}
uniformly for $\xi$ on every compact subset of the interval $(0,1)$.
Since $f'(u) < 0$, we estimate for $u_1 > \xi$,
$$\int_{\xi}^{u_1} {f'(u) \over \sqrt{u - \xi } } d u \leq \int_{\xi}^{u_1} f'(u) d u 
= f(u_1) - f(\xi) \ .$$
As $u_1 \rightarrow 1$, the right hand side goes to $- \infty$ uniformly for $\xi$ 
on every compact subset of the interval $(0,1)$ because $f(u_1) \rightarrow - \infty$
in view of (\ref{cf1}) and $f(u)$ is bounded on every compact subset.
This proves the limit (\ref{f'}).

To calculate the limit (\ref{F'}), we turn to the following lemma on the uniform
convergence version of L'Hospital's rule. The proof of Lemma \ref{L'Hospital}
is simply a modification of the standard proof of the classical L'Hospital's rule.

\begin{lem}
\label{L'Hospital}
Let $g_1(x,y)$ and $g_2(x,y)$ be defined on $(0, 1) \times Y$, where $Y$ is a subset of
$\Re$. If $g_1(x,y)$ and $g_2(x,y)$ satisfy the conditions
\begin{enumerate}

\item for each $x \in (0,1)$, $g_1(x,y)$ and $g_2(x,y)$ are uniformly bounded for $y$ on $Y$,

\item the partial derivatives $g_{1x}(x,y)$ and $g_{2x}(x,y)$ exist for each $(x,y) \in (0, 1)
\times Y$,

\item $\lim_{x \rightarrow 1} g_2(x,y) = \infty$ uniformly for $y$ on $Y$,

\item $\lim_{x \rightarrow 1} [g_{1x}(x,y) / g_{2x}(x,y)] = L(y)$ uniformly for $y$ on $Y$,

\item $L(y)$ is uniformly bounded on $Y$.

\end{enumerate}
then 
$\lim_{x \rightarrow 1} [g_1(x,y) / g_2(x,y)] = L(y)$
uniformly for $y$ on $Y$.
\end{lem}

Using formulae (\ref{fF2}) and (\ref{fF2'}), we apply Lemma \ref{L'Hospital} twice to calculate the limit
\begin{eqnarray*}
\lim_{u_1 \rightarrow 1} { {\p F_0(\xi, u_1) \over \p \xi} \over F_0(\xi, u_1) }
&=& \lim_{u_1 \rightarrow 1} {1 \over u_1 - \xi} { \int_{\xi}^{u_1}   
{ f''(u) \over \sqrt{u - \xi} } (u_1 - u) d u \over \int_{\xi}^{u_1} {f'(u)
\over \sqrt{u - \xi } } d u }  \\
&=& {1 \over 1 - \xi} \lim_{u_1 \rightarrow 1} {\int_{\xi}^{u_1}
{f''(u) \over \sqrt{u - \xi} } d u \over {f'(u_1) \over \sqrt{u_1 - \xi} }}  \\
&=& {1 \over \sqrt{1 - \xi} } \lim_{u_1 \rightarrow 1} { {f''(u_1) \over \sqrt{u_1 - \xi}}
\over f''(u_1)}  \\
&=& {1 \over 1 - \xi} \ ,  
\end{eqnarray*}
uniformly for $\xi$ on every compact subset of $(0,1)$.
Here, the limit (\ref{f'}) and the second limit of (\ref{cf3}) guarantee that Lemma \ref{L'Hospital}
applies.
This proves the limit (\ref{F'}).

The proof of Lemma \ref{trailing} is now complete. 

We now use Lemma \ref{trailing} to study the minimum problem (\ref{min}). Because
of (\ref{F}), $[- F_0(\xi, u_1) / U_0(\xi, u_1)]$ goes to $+ \infty$ as
$\xi \rightarrow 0, 1$. Hence, the minimum is reached somewhere in 
the interval $(0, 1)$.

We next show that the minimum problem (\ref{min}) has only one critical point when $u_1$ is close
to $1$. 
To see this, we calculate the second derivatives of $[- F_0 / U_0]$
at any of its critical points
\begin{equation}
\label{''} {\p^2 \over \p \xi^2} [- {F_0(\xi, u_1) \over U_0(\xi, u_1)}] =
- {F_{0 \xi \xi} U_0 - F_0 U_{0 \xi \xi} \over U_0^2} \ .
\end{equation}
Since $U_{0 \xi \xi} \equiv 0$ in view of (\ref{fU}), it follows from inequality
(\ref{F_0''}) that the right hand side is positive if $u_1$ is close to $1$.
Hence, all the critical points of $[- F_0 / U_0]$ must be local
minimizing points.
For any two minimizing points, some point between them
must be a local maximizing point; an impossibility. This proves
the uniqueness of the critical point and hence of the minimizing point
in the minimum problem (\ref{min}) when $u_1$ is close to $1$.

\noindent
{\bf Remark:} \ {\em This is the first place where $m=2$ in (\ref{am}) is important.
It is because of $m=2$ that we have $U_{0 \xi \xi} \equiv 0$ which makes
it possible to conclude from formula (\ref{''}) the uniqueness of the critical point
in the the minimum problem (\ref{min}).}  

We also want to determine the asymptotics of the critical point when $t$ is large.
To do this, we write 
\begin{equation}
\label{F0/U0}
{\p \over \p \xi} [- {F_0(\xi, u_1) \over U_0(\xi, u_1)}] =
- {F_0 \over U_0}[ {{\p F_0 \over \p \xi} \over F_0} -
{{\p U_0 \over \p \xi} \over U_0} ] \ .
\end{equation}
Since $F_0$ and $U_0$ are non-zero when $0 < \xi < 1$ and $0 < u_1 < 1$,
$[- F_0/U_0 ]$ has a unique critical point within the interval $(0,1)$ 
if and only if the quantity in the parenthesis on the right hand side does so.
We now consider a 
compact set $[\epsilon_0, 1 - \epsilon_0]$, where $\epsilon_0$ is a small positive
number. According to (\ref{F'}),  
the quantity in the parenthesis of (\ref{F0/U0}) has the uniform limit
\begin{equation}
\label{U0}
{1 \over 1 - \xi} - {{\p U_0(\xi, 1) \over \p \xi} \over U_0(\xi, 1)}
\end{equation}
on this compact set as $u_1 \rightarrow 1$. In view of formula (\ref{fU}) for $U_0$, we deduce that
the function (\ref{U0}) has only one zero $\xi = 1/4$ in the interval $0 < \xi < 1$, 
counting multiplicities.
Hence, the quantity 
in the parenthesis of
(\ref{F0/U0}) will have a zero near $\xi = 1/4$ when $u_1$ is close to $1$. This zero is exactly the unique 
critical point of the minimum problem (\ref{min}).

We summarize the above result as a lemma.
\begin{lem}
\label{trailing2}
The minimum problem
(\ref{min}) has one and only one critical point  
when $u_1$ is close to $1$. Furthermore, the critical point approaches $1/4$ as
$u_1$ goes to $1$ .
\end{lem}

We are now ready to solve equations (\ref{teg10}-\ref{teg30}). By Lemma \ref{trailing2}, 
equation (\ref{teg20'}) has
a unique solution $\xi = u_3$ for each $u_1$ close to $1$.  
We then use (\ref{teg10'}) to determine $t$. Hence, equations (\ref{teg10'}-\ref{teg20'})
can be solved to give $u_3$ and $t$ as functions of $u_1$.   

We then calculate the Jacobian of equations (\ref{teg10}-\ref{teg20}) with respect to
$u_1$ and $u_3$ on the solution of (\ref{teg10'}-\ref{teg20'}):
\begin{eqnarray}
{\p W_1 \over \p u_1} &=& {\Phi_0(u_1, u_1) \over 2(u_1 - u_3)} >  0 \ , \label{W11} \\
{\p W_1 \over \p u_3} &=& 0 \ , \label{W13} \\
{\p W_2 \over \p u_1} &=& {\Phi_0(u_1, u_1) \over 2(u_1 - u_3)^2} >  0 \ , \label{W21} \\
{\p W_2 \over \p u_3} &=& - {\p^2 \over \p \xi^2} F_0(\xi, u_1)|_{\xi = u_3}  > 0 \label{W23} \ .
\end{eqnarray}

To prove the equality of (\ref{W11}), we use the special case of identity (\ref{Phiu})
$$2(\xi - u_1) {\p \over \p u_1} \Phi_0(\xi, u_1) = \Phi_0(\xi, u_1) - \Phi_0(u_1, u_1) \ .$$
This together with equation (\ref{teg10}) proves the equality of (\ref{W11}). 

Equation (\ref{W13}) is the result of (\ref{teg20}).

The equality of (\ref{W21}) follows from (\ref{ph2}), (\ref{teg20}) and (\ref{W11}).

The equality of (\ref{W23}) is the result of 
the split (\ref{split}) and $U_{0 \xi \xi} \equiv 0$.

The sign in (\ref{W23}) comes from (\ref{F_0''}).

To determine the signs in (\ref{W11}) and (\ref{W21}), we split $\Phi_0(u_1, u_1)$
according to (\ref{split}),
\begin{eqnarray*}
\Phi_0(u_1,u_1) &=& - 30 t U_0(u_1,u_1) - F_0(u_1, u_1)  \\
                &=& U_0(u_1, u_1) [ {F_0(u_3, u_1) \over U_0(u_3, u_1)} - 
{F_0(u_1, u_1) \over U_0(u_1, u_1)} ] > 0 \ .
\end{eqnarray*}
Here, we have used (\ref{teg10'}) in the second equality. The inequality follows
from the fact that $u_3$ is the only minimizing point of the minimum problem
(\ref{min}).

It now follows from the Implicit Function Theorem that equations (\ref{teg10}-\ref{teg20})
can be solved to give $u_1$ and $u_3$ as functions of $t$,
respectively, for $t$ larger than a finite time. We have therefore established
the following theorem.

\begin{theo}
\label{trailing3}
Under the conditions (\ref{cf1}) and (\ref{cf2}), there exists a constant $T^- > 0$ such that
for $t > T^-$, equations (\ref{teg10}-\ref{teg30})  
have a unique solution $u_1^-(t)$, $u_3^-(t)$ and $x^-(t)$.
%with $u_3^-(t)$ being the
%unique minimizer of (\ref{min}) for each $u_1=u_1^-(t)$.
Function $u_1^-(t)$ is an increasing function of $t$.
Furthermore,
$$\lim_{t \rightarrow + \infty} u_1^-(t) = 1 \ , ~~~~~
\lim_{t \rightarrow + \infty} u_3^-(t) = {1 \over 4} \ . $$
\end{theo}

\medskip

\noindent
{\bf $\S$ 4.2 \quad The leading edge}

We now turn to the leading edge at which $u_1 = u_2$. It is the head of a single phase solution (see Figure 1.). 
It is governed by equations (\ref{leg1}-\ref{leg3})
for $g=0$:   
\begin{eqnarray}
W_3(u_1, u_3, t) =: \Phi_0(u_1, u_3)
&=& 0 \ , \label{leg10} \\
W_4(u_1, u_3, t) =: \int_{u_3}^{u_1} \Phi_0(\xi, u_3) \sqrt{\xi - u_3} d \xi &=& 0 \ ,
\label{leg20} \\
x - 30 t u_3^2 - f(u_3) &=& 0 \ . \label{leg30}
\end{eqnarray}

We again use the split $\Phi_0(\xi, u_3) = - 30 t U_0(\xi, u_3) - F_0(\xi, u_3)$,
which is similar to (\ref{split}),
to rewrite (\ref{leg10}-\ref{leg20}) as
\begin{eqnarray}
{\int_{u_3}^{u_1} F_0(u, u_3) \sqrt{u - u_3} d u  \over 
\int_{u_3}^{u_1} U_0(u, u_3) \sqrt{u - u_3} d u} + 30 t &=& 0 \ , \label{leg10'} \\
{\p \over \p u_1} \left[ {\int_{u_3}^{u_1} F_0(u, u_3) \sqrt{u - u_3} d u  \over     
\int_{u_3}^{u_1} U_0(u, u_3) \sqrt{u - u_3} d u} + 30 t \right ] &=& 0 \ . \label{leg20'}
\end{eqnarray}

We first simplify equations (\ref{leg10'}-\ref{leg20'}). 

\begin{lem}
\label{FG}
$${1 \over 2 (u_1 - u_3)^{3 \over 2}} \int_{u_3}^{u_1} F_0(u, u_3) \sqrt{u - u_3} d u
= {1 \over 4 \sqrt{2}} \int_{-1}^1 f'({1 - \mu \over 2} u_3 + {1 + \mu \over 2} u_1)
\sqrt{1 - \mu} d \mu \ .$$
\end{lem}
Proof: 
\begin{eqnarray*}
{\p \over \p u_1} \int_{u_3}^{u_1} F_0(u, u_3) \sqrt{u - u_3} d u
&=& F_0(u_1, u_3) \sqrt{u_1 - u_3} \\
&=& {1 \over 2} \int_{u_3}^{u_1} {f'(u) \over \sqrt{u_1 - u}} d u \\
&=& {\p \over \p u_1} \int_{u_3}^{u_1} f'(u) \sqrt{u_1 - u} d u \\
&=& {\p \over \p u_1} [{(u_1 - u_3)^{3 \over 2} \over 2 \sqrt{2}} 
\int_{-1}^1 f'({1 - \mu \over 2} u_3 + {1 + \mu \over 2} u_1)
\sqrt{1 - \mu} d \mu] \ , 
\end{eqnarray*}
which proves the lemma.

Let 
\begin{eqnarray}
G_0(\xi, u_3) &=& {1 \over 4 \sqrt{2}} \int_{-1}^1 f'({1 - \mu \over 2} u_3 + {1 + \mu \over 2} \xi)
\sqrt{1 - \mu} d \mu \ , \label{fG} \\
V_0(\xi, u_3) &=& {1 \over 4 \sqrt{2}} \int_{-1}^1 2 ({1 - \mu \over 2} u_3 + {1 + \mu \over 2} \xi)
\sqrt{1 - \mu} d \mu = {4 \over 15 } \xi + {2 \over 5} u_3 \ . \label{fV} 
\end{eqnarray}

We use Lemma \ref{FG} to rewrite equations (\ref{leg10'}-\ref{leg20'}) as
\begin{eqnarray}
{G_0(\xi, u_3) \over V_0(\xi, u_3)} + 30 t &=& 0
\ , \label{leg10''} \\
{\p \over \p \xi} [{G_0(\xi, u_3) \over V_0(\xi, u_3)} + 30 t]_{\xi = u_1}
&=& 0  \ . \label{leg20''}
\end{eqnarray}
As before, equation (\ref{leg20''}) motivates us to consider another auxiliary minimum problem
\begin{equation}
\label{min2}
Min_{0 < \xi < 1} \left[ - {G_0(\xi, u_3) \over V_0(\xi, u_3)} \right] 
\end{equation}
for each $u_3$ close to $0$. 

\begin{lem}
\label{leading}

\begin{enumerate}

\item For each $u_3 \in (0, 1)$, 
\begin{equation}
\label{G}
lim_{\xi \rightarrow 1} {\p \over \p \xi} 
\left[ - {G_0(\xi, u_3) \over V_0(\xi, u_3)} \right] = + \infty \ .
\end{equation}

\item For any small $\epsilon_1 > 0$, there exists a $\delta^* > 0$ such that
\begin{equation}
\label{G'}
{\p \over \p \xi}
\left[ - {G_0(\xi, u_3) \over V_0(\xi, u_3)} \right] < 0 
\end{equation}
for $0 < \xi < 1 - \epsilon_1$ and $0 < u_3 < \delta^*$.

\item  
\begin{equation}
\label{G'2}
{\p^2 \over \p \xi^2} G_0(\xi, u_3) < 0 \quad uniformly \ for \ 0 < \xi < 1
\end{equation}
when $u_3$ is close to $0$.

\end{enumerate}

\end{lem}

Proof: We omit the proof of (\ref{G'2}) since it is similar to the proof of
(\ref{F_0''}).

To prove (\ref{G}), we write
\begin{equation}
\label{G/V}
{\p \over \p \xi} [- {G_0(\xi, u_3) \over V_0(\xi, u_3)}] =
-  {{\p G_0 \over \p \xi} V_0 - G_0
{\p V_0 \over \p \xi} \over V_0^2} \ .
\end{equation}

We use formula (\ref{fG}) to calculate $\p_{\xi} G_0$ and integrate by parts to obtain
\begin{equation}
\label{GF0}
{\p G_0 \over \p \xi} =  {F_0(\xi, u_3) - 3 G_0(\xi, u_3) \over 2(\xi - u_3)} \ ,
\end{equation}
where $F_0$ is previously defined by (\ref{fF}). 

Using (\ref{fG}) again, we write 
\begin{equation*}
G_0(\xi, u_3) = {1 \over 2(\xi - u_3)^{3 \over 2}} \int_{u_3}^{\xi} f'(u) \sqrt{\xi -u}
\ d u \ .
\end{equation*}
Since the integral is a decreasing function of $\xi$ because $f'(u) < 0$, $G_0(\xi, u_3)$   
goes to either a finite negative number or $- \infty$ as $\xi \rightarrow 1$.

For those initial functions $f(u)$ such that $G_0(\xi, u_3)$ converges as $\xi \rightarrow 1$, 
the right hand side 
of (\ref{GF0}) goes to $- \infty$ because $F_0(\xi, u_3)$ does so by Lemma \ref{trailing};
so $G_{0 \xi}(\xi, u_3) \rightarrow - \infty$. This together with (\ref{G/V}) gives the limit   
(\ref{G}).

For those initial functions $f(u)$ such that $G_0(\xi, u_3) \rightarrow - \infty$ as $\xi \rightarrow 1$, we estimate 
$${F_0(\xi, u_3) \over G_0(\xi, u_3)} = { \int_{u_3}^{\xi} {f'(u) \over \sqrt{\xi - u}} d u \over (\xi - u_3)
\int_{u_3}^{\xi} f'(u) \sqrt{\xi - u} d u} \geq { \int_{u_3}^{\xi} {f'(u) \over \sqrt{1 - u}} d u \over
\int_{u_3}^{\xi} f'(u) \sqrt{1 - u} d u} \ .$$
Using L'Hospital's rule, it is easy to see that the right hand side goes to $+ \infty$ as
$\xi \rightarrow 1$.  This limit, when combined with (\ref{GF0}), gives 
$$\lim_{\xi \rightarrow 1} {{\p G_0(\xi, u_3) \over \p \xi} \over G_0(\xi, u_3)}
= + \infty \ , $$
which, in view of (\ref{G/V}), proves (\ref{G}).

The proof of the first part of Lemma \ref{leading} is completed.

To prove (\ref{G'}), we use (\ref{G/V}) again.
Since $V_0$ and $\p_{\xi} V_0$ are positive functions and $G_0$ is negative, 
it suffices to show that 
\begin{equation}
\label{G0}
{\p G_0(\xi, u_3) \over \p \xi} > 0
\end{equation}
for $0 < \xi < 1 - \epsilon_1$ and $0 < u_3 < \delta$.

In view of the third limit of (\ref{cf3}), there exists a $\delta_2 > 0$ such that
\begin{equation}
\label{f''}
f''(u) > 0 \quad \mbox{for $0 < u < \delta_2$} \ .
\end{equation}
It then follows from formula (\ref{fG}) for $G_0$ that
inequality (\ref{G0}) is true for $0 < \xi < \delta_2$ and $0 < u_3 < \delta_2$.

For $\delta_2 \leq \xi < 1 - \epsilon_1$ and $0 < u_3 < \delta_2 / 2$, we use formula (\ref{fG}) again to write
\begin{eqnarray*}
{\p G_0(\xi, u_3) \over \p \xi} &=& {1 \over 2(\xi - u_3)^{5 \over 2}} \int_{u_3}^{\xi} f''(u) \sqrt{\xi -u}
(u - u_3) d u  \no \\   
&=& {1 \over 2(\xi - u_3)^{5 \over 2}} \int_{u_3}^{\delta_2 \over 2} f''(u) \sqrt{\xi -u}
(u - u_3) d u \\
&& + {1 \over 2(\xi - u_3)^{5 \over 2}} \int_{\delta_2 \over 2}^{\xi} f''(u) \sqrt{\xi -u} (u - u_3) d u \ . 
\end{eqnarray*}
The second term is bounded from below by 
$$ - \ {2^{3 \over 2} \over \delta_2^{5 \over 2}} \int_{\delta_2 \over 2}^{1 - \epsilon_1} |f''(u)|
d u \ ,$$
which is a constant.
Since $f'' > 0$ for $0 < u < \delta_2$, the first term is bigger than 
$$ {\sqrt{\delta_2} \over 2 \sqrt{2}}  \int_{u_3}^{\delta_2 \over 2} f''(u) (u - u_3) d u 
= {\sqrt{\delta_2} \over 2 \sqrt{2}} [({\delta_2 \over 2} - u_3)f'({\delta_2 \over 2}) - f({\delta_2 \over 2}) +
f(u_3)] \ ,$$
which goes to $+ \infty$ as $u_3 \rightarrow 0$ because of (\ref{cf1}).
We have therefore proved (\ref{G0}) and hence (\ref{G'}).
 
The proof of Lemma \ref{leading} is complete.
 
We now use Lemma \ref{leading} to study the minimization problem (\ref{min2}).
For each small $u_3$, the function $[- G_0(\xi, u_3) / V_0(\xi, u_3)]$ is decreasing when $\xi < 1 - \epsilon_1$
in view of (\ref{G'}) and tending to $+ \infty$ as $\xi \rightarrow 1$ in view of (\ref{G}).
Hence, the minimum is reached somewhere in the
interval $(0, 1)$.

We next show that the minimum problem (\ref{min2}) has only one critical point when $u_3$ is close
to $0$.
To see this, we calculate the second derivatives of $[- G_0 / V_0]$
at any of its critical points
\begin{equation}
\label{GV2}
{\p^2 \over \p \xi^2} [- {G_0(\xi, u_3) \over V_0(\xi, u_3)}] =
- {G_{0 \xi \xi} V_0 - G_0 V_{0 \xi \xi} \over V_0^2} \ .
\end{equation}
Since $V_{0 \xi \xi} \equiv 0$ from (\ref{fV}), the right hand side of
(\ref{GV2}) is positive 
if $u_3$ is close to $0$, in view
of (\ref{G'2}). Hence, all the critical points of $[- G_0 / V_0]$ must be local
minimizing points.
For any two minimizing points, some point between them
must be a local maximizing point; an impossibility. This proves
the uniqueness of the critical point and hence of the minimizing point
in the minimum problem (\ref{min2}) when $u_3$ is close to $0$.

\noindent
{\bf Remark:} \ {\em This is the second place where $m=2$ in (\ref{am}) plays an important role.
Since $m=2$, $V$ of (\ref{fV}) is a linear function. It then becomes possible to conclude from 
formula (\ref{GV2}) that the minimum problem (\ref{min2}) has a unique critical point.}

The critical point approaches $1$ as $u_3$ goes to $0$. This follows from
(\ref{G'}).

We summarize the above result as a lemma.
\begin{lem}
\label{leading2}
The minimum problem
(\ref{min2}) has one and only one critical point and hence minimizing point
when $u_3$ is close to $0$. Furthermore, the critical point tends to $1$ as
$u_3$ goes to $0$.
\end{lem}

We are now ready to solve equations (\ref{leg10}-\ref{leg30}). By Lemma \ref{leading2},
equation (\ref{leg20''}) has
a unique solution $\xi = u_1$ for each $u_3$ close to $0$.
We then use (\ref{leg10''}) to determine $t$. Hence, equations (\ref{leg10''}-\ref{leg20''})
or equivalently equations (\ref{leg10}-\ref{leg30}) can be solved to give $u_3$ and $t$ as 
functions of $u_1$.

We then calculate the Jacobian of equations (\ref{leg10}-\ref{leg20}) with respect to
$u_1$ and $u_3$ on the solution of (\ref{leg10''}-\ref{leg20''}):
\begin{eqnarray}
{\p W_3 \over \p u_1} &=& {\p \Phi_0(\xi, u_3) \over \p \xi}|_{\xi = u_1} >  0 \ , \label{W31} \\
{\p W_3 \over \p u_3} &=& - {\Phi_0(u_3, u_3) \over 2(u_1 - u_3)} < 0  \ , \label{W33} \\
{\p W_4 \over \p u_1} &=& 0 \ , \label{W41} \\
{\p W_4 \over \p u_3} &=& - \Phi_0(u_3, u_3) \sqrt{u_1 - u_3} < 0 \ . \label{W43}
\end{eqnarray}

The equalities in (\ref{W33}) and (\ref{W43}) follow from the related identity (\ref{Phiu})
for $\Phi_0(\xi, u_3)$.

Equation (\ref{W41}) is the result of (\ref{teg20}).

The signs in (\ref{W33}) and (\ref{W43}) are determined as follows.
Since the minimum problem (\ref{min2}) has only one minimizing point at 
$\xi = u_1$, we must have 
$$ - {G_0(u_3, u_3) \over V_0(u_3, u_3)} > - {G_0(u_1, u_3) \over V_0(u_1, u_3)} \ .$$
The right hand side is equal to $30 t$ in view of (\ref{leg10''}). Hence,
$- 30 t V_0(u_3, u_3) - G_0(u_3, u_3) > 0 \ ,$ which by formulae (\ref{fG}-\ref{fV}), reduces to
$$- 60 t  u_3 - f'(u_3) > 0 \ .$$
In view of the boundary condition (\ref{ph3}) for $\Phi_0$, this is exactly $\Phi_0(u_3, u_3) > 0$, 
which proves the inequalities in (\ref{W33}) and (\ref{W43}).

To determine the sign in (\ref{W31}), since $\xi = u_1$ is the minimizing point in the minimum
problem (\ref{min2}), we have 
$$ {\p^2 \over \p \xi^2} [ - {G_0(\xi, u_3) \over V_0(\xi, u_3)} ] |_{\xi = u_1} > 0 \ ,$$
which, in view of Lemma \ref{FG}, is equivalent to
$${\p^2 \over \p \xi^2} \left[ - {\int_{u_3}^{\xi} F_0(u, u_3) \sqrt{u - u_3} d u  \over
\int_{u_3}^{\xi} U_0(u, u_3) \sqrt{u - u_3} d u} \right ]|_{\xi = u_1} > 0 \ . $$
This together with (\ref{leg10'}-\ref{leg20'})) gives 
$${\p^2 \over \p \xi^2} \int_{u_3}^{\xi} [- 30 t U_0(u, u_3) - F_0(u, u_3)] \sqrt{u - u_3} d u
> 0 $$ at $\xi = u_1$. Since $\Phi_0 = - 30 t U_0 - F_0$, the last inequality together 
with (\ref{leg10}) implies that
$${\p \Phi_0(\xi, u_3) \over \p \xi}|_{\xi=u_1} > 0 \ ,$$
which proves the inequality in (\ref{W31}).

We have therefore established the following theorem.

\begin{theo}
\label{leading3}
Under the conditions (\ref{cf1}) and (\ref{cf2}), there exists a constant $T^+ > 0$ such that
equations (\ref{leg10}-\ref{leg30})
have a unique solution $u_1^+(t)$, $u_3^+(t)$ and $x^+(t)$ for $t > T^+$.
Functions $u_1^+(t)$ and $u_3^+(t)$ are increasing and decreasing functions of $t$, respectively.
They have the limits:
$$\lim_{t \rightarrow + \infty} u_1^+(t) = 1 \ , ~~~~~
\lim_{t \rightarrow + \infty} u_3^+(t) = 0 \ . $$
Furthermore, 
$$ 1 - \delta < u_1^+(t) < 1 \ , \quad 0 < u_3^+(t) < \delta_2 \quad \mbox{for $t > T^+$} \ ,$$
where $\delta$ and $\delta_2$ are defined in (\ref{F_0''}) and (\ref{f''}), respectively.
\end{theo}

The second part of Theorem \ref{leading3} is only for technical purpose. It will be needed
in the next section.

\bigskip

\refstepcounter{section}
\begin{center}
{\bf $\S$ 5 \quad Analysis of Zero and Single Phase Solutions}
\end{center}

In this section, we shall show that the hodograph transform (\ref{P2})
$g=0, 1$, can be inverted to give solutions $u_1(x,t)$, $\cdots$, $u_{2g+1}(x,t)$
for $- \infty < x < \infty$ and $t > max \{ T^-, T^+ \}$, where $T^-$ and $T^+$
are given in Theorems \ref{trailing3} and \ref{leading3}, respectively. We will also verify 
inequalities (\ref{constraint}-\ref{constraint2}). By theorem \ref{main}, the 
function $\psi$ of (\ref{psi}) will be the minimizer of (\ref{mini}).

For each $t > max \{ T^-, T^+ \}$, we split the space into three pieces $(- \infty, x^-(t)]$,
$[x^-(t), x^+(t)]$ and $[x^+(t), + \infty)$. Here, $x^-(t)$ and $x^+(t)$ are given
in Theorems  \ref{trailing3} and \ref{leading3}, respectively.  

\noindent
{\bf $\S$ 5.1 \quad Zero phases:  $x \leq x^-(t)$ and $x \geq x^+(t)$}

We shall focus on the zero phase over the interval $x \geq x^+(t)$.
The other zero phase over $x \leq x^-(t)$ can be handled in the same way.

For $x \geq x^+(t)$, we shall use the auxiliary minimum problem (\ref{min2}) to
show that the minimizer of the minimization problem (\ref{mini})
has a compact support $[0, \sqrt{u_3(x,t)}]$. The function $u_3(x,t)$ is governed by
equation (\ref{P2}), which, when $g=0$, becomes  
\begin{equation}
\label{5Burgers'}
 x - 30 t u_3^2 - f(u_3) = 0 \ . 
\end{equation}

For each $0 < u_3 < u_3^+(t)$ where $u_3^+(t)$ is given in Theorem \ref{leading3}, since 
$u_3^+(t)$ is a continuous function of $t$ and $u_3^+(t) \rightarrow 0$ as $t \rightarrow
+ \infty$,
there is a $t^* > t$ such that $u_3 = u_3^+(t^*)$. In view of the minimum problem
(\ref{min2}), we have for $0 < \xi < 1$, 
$$ - {G_0(\xi, u_3) \over V_0(\xi, u_3)} \geq - {G_0(u_1^+(t^*), u_3) \over V_0(u_1^+(t^*), u_3)} 
= 30 t^* > 30 t \ ,$$
where the equality follows from equation (\ref{leg10''}). Hence,  
\begin{equation}
\label{B}
- 30 t V_0(\xi, u_3) - G_0(\xi, u_3) > 0 
\end{equation}
for all $0 < \xi < 1$.  In particular, when $\xi = u_3$, inequality (\ref{B})
reduces to $- 60 t u_3 - f'(u_3) > 0$ 
in view of formulae (\ref{fG}) and (\ref{fV}). This shows that equation (\ref{5Burgers'})
can be inverted to give $u_3$ as a decreasing function of $x$ for $x > x^+(t)$.

We next verify inequalities (\ref{constraint}-\ref{constraint2}), which now take the form
\begin{eqnarray}
Re\{\sqrt{-1} \sqrt{\xi - u_3}\}\Phi_0(\xi,
u_3) & < & 0 \quad
\mbox{for $0 < \xi < u_3$} \ , \label{constraint'} \\
\int_{u_3}^{\xi}\sqrt{u - u_3} \Phi_0(u, u_3)
d u  & > &  0 \quad \mbox{for $\xi > u_3$} \ .
\label{constraint2'}
\end{eqnarray}

Inequality (\ref{constraint2'}) follows easily from (\ref{B}) and Lemma \ref{FG}.

To prove (\ref{constraint'}), we use $u_3 < u_3^+(t)$ again. Since $u_3^+(t) < \delta_2$ according 
to Theorem \ref{leading3}, we have $u_3 < \delta_2$. We then use (\ref{f''}) in formula
(\ref{fF}) for $F_0$ to obtain that
$${\p F_0(\xi, u_3) \over \p \xi} > 0$$
for $\xi \leq u_3 < \delta_2$. Hence,
$$ {\p \Phi_0(\xi, u_3) \over \p \xi} = - 30 t {\p U_0(\xi, u_3) \over \p \xi} 
- {\p F_0(\xi, u_3) \over \p \xi} < 0 $$
for all $\xi \leq u_3$. This together with $\Phi_0(u_3, u_3) = 
- 60 t u_3 - f'(u_3) > 0$ proves $\Phi_0(\xi, u_3) > 0$ for
$\xi \leq u_3$ and hence inequality (\ref{constraint'}).

\begin{theo}
\label{Burgers2}
Under the conditions (\ref{cf1}) and (\ref{cf2}), for each $t > max \{ T^-, T^+ \}$, the
minimizer of the minimization problem (\ref{mini}) is supported on a single interval
when $x \leq x^-(t)$ and $x \geq x^+(t)$.
\end{theo}
 
\noindent
{\bf $\S$ 5.2 \quad Single phase:  $x^-(t) < x < x^+(t)$}

For $t > max \{ T^-, T^+ \}$ and $x^-(t) < x < x^+(t)$, we shall 
show that the minimizer of (\ref{mini}) 
is supported on two disjoint intervals $[0, \sqrt{u_3(x,t)}]$ and $[\sqrt{u_2(x,t)}, 
\sqrt{u_1(x,t)}]$, where $u_1(x,t)$, $u_2(x,t)$ and $u_3(x,t)$ are governed by
equations (\ref{P2}) when $g=1$
\begin{equation}
\label{P2'}
P(u_1, u_1, u_2, u_3) = 0 \ , ~~ P(u_2, u_1, u_2, u_3) = 0 \ , ~~ P(u_3, u_1, u_2, u_3) = 0 \ . 
\end{equation}

We will first solve equations (\ref{P2'}) near the leading edge $x=x^+(t)$.
We will then extend the solution over the whole interval $x^-(t) < x < x^+(t)$. 

The first two equations of (\ref{P2'}) are degenerate near the leading edge
at which $u_1 = u_2$. We replace them by 
\begin{eqnarray}
W_5(u_1, u_2, u_3, t) =: {\p q_{1,1} \over \p u_1} + {\p q_{1,1} \over \p u_2}
+ {\p q_{1,1} \over \p u_3} &=& 0  , \label{l1} \\
W_6(u_1, u_2, u_3, t) =: \int_{u_3}^{u_2} \sqrt{(u_1 - \xi)(u_2 - \xi)(\xi - u_3)}
\Phi_1(\xi, u_1, u_2, u_3) d \xi &=& 0   , \label{l2}
\end{eqnarray}
where $\Phi_1$ and $q_{1,1}$ are given by (\ref{ph1}-\ref{ph3}) and (\ref{q1}-\ref{q2}), respectively.
It can be shown that equations (\ref{l1}) and (\ref{l2}) together with $P(u_3, \vu) = 0$
are equivalent
to equations (\ref{P2'}) when $u_1 > u_2 > u_3$ \cite{gra1}.

It is also known that equations (\ref{l1}) and (\ref{l2}) transform into
equations (\ref{leg10}-\ref{leg20}) when $u_1=u_2$ and
that equation $P(u_3, \vu) = 0$ into (\ref{leg30}) \cite{gra1}.

For each fixed $t$, we now solve equations (\ref{l1}) and (\ref{l2}) for $u_1$ and $u_3$ in terms of
$u_2$ near the leading edge. We use the identities of Lemma \ref{gidentity} to calculate the derivatives
of $W_5$ and $W_6$ at the point $(u_1^+(t), u_1^+(t), u_3^+(t))$:
\begin{eqnarray}
{\p W_5 \over \p u_1} &=& {\p W_5 \over \p u_2} = {1 \over 2} {\p \Phi_0(\xi, u_3) \over \p \xi}|_{\xi = u_1} >  0 \ , \label{W51} \\
{\p W_5 \over \p u_3} &=& {\Phi_0(u_1, u_3) - \Phi_0(u_3, u_3) \over 2(u_1 - u_3)}
 = - {\Phi_0(u_3, u_3) \over 2(u_1 - u_3)} < 0  \ , \label{W53} \\
{\p W_6 \over \p u_1} &=& {\p W_6 \over \p u_2} =  {1 \over 2} \int_{u_3}^{u_2} \sqrt{\xi - u_3}
[2(u_1 - \xi) {\p \Phi_1(\xi, \vu) \over \p u_1} + \Phi_1(\xi, \vu) ] d \xi \no \\
&=& {(u_1 - u_3)^{3 \over 2} \over 3} \Phi_1(u_1, \vu) = {(u_1 - u_3)^{3 \over 2} \over 3} {\p \Phi_0(\xi, u_3)
\over \p \xi}|_{\xi = u_1} > 0 \ , \label{W61} \\
{\p W_6 \over \p u_3} &=&  - \int_{u_3}^{u_2} {u_1 - \xi \over 2 \sqrt{\xi - u_3} }
[ 2 (u_3 - \xi) {\p \Phi_1(\xi, \vu) \over \p u_3} + \Phi_1(\xi, \vu) ] d \xi \no \\
&=& - \int_{u_3}^{u_2} {u_1 - \xi \over 2 \sqrt{\xi - u_3} } \Phi_1(u_3, \vu) d \xi \no \\
&=& - {2 (u_1 - u_3)^{3 \over 2} \over 3} [ {\Phi_0(u_3, u_3) - \Phi_0(u_1, u_3) \over u_3 - u_1} ] \no \\ 
&=&  {2 \sqrt{u_1 - u_3} \over 3 } \Phi_0(u_3, u_3) > 0 \ , \label{W63}
\end{eqnarray}
where $\vec{u} = (u_1, u_2, u_3)$.

The second equality of (\ref{W51}) follows from identity (\ref{lim3}).  

The first equality of (\ref{W53}) is a consequence of (\ref{Phiu}) and (\ref{lim3}). The second
equality follows from equation (\ref{leg10}).  

The third and last equalities of (\ref{W61}) follows from identities (\ref{Phiu}) and (\ref{lim2}),
respectively.

The second, third and last equalities of (\ref{W63}) follows from (\ref{Phiu}), (\ref{lim1}) and
(\ref{leg10}), respectively.   
 
All the signs are easily determined according to (\ref{W31}-\ref{W43}).

Therefore, equations (\ref{l1}) and (\ref{l2}) can be inverted to give $u_1$ and $u_3$ as 
functions of $u_2$ when $u_2$ is close to and less than $u_1^+(t)$. Moreover,
$u_1$ decreases as $u_2$ increases. We then use the last equation of (\ref{P2'}) to give
$x$ as a function of $u_2$. We summarize these result in the lemma.

\begin{lem}
\label{leadingn}
For each $t > T$, equations (\ref{P2'}) can be inverted to give $x$, $u_1$ and $u_3$
as functions of $u_2$ when $u_2$ is close to and less than $u_1^+(t)$.
\end{lem}

Having solved equations (\ref{P2'}) for $x$, $u_1$ and $u_3$ as functions of $u_2$
near the leading edge $u_2 = u_2^+(t)$, we now extend the solution by decreasing $u_2$.
To calculate the Jacobian of (\ref{P2'}), we turn to Lemma \ref{jaco}, which says
that the zeros of the Jacobian are determined by the $\xi$-zeros of $\Phi_1(\xi, u_1,u_2,u_3)$.
We again split
\begin{equation}
\label{split3}
\Phi_1(\xi, u_1,u_2,u_3) = - 30 t U_1(\xi, u_1,u_2,u_3) - F_1(\xi, u_1,u_2,u_3) \ ,
\end{equation}
where $U_1$ and $F_1$ satisfy the Euler-Poisson-Darboux equations (\ref{ph1}-\ref{ph3})
with the boundary values $4/3$ and $2 f''(u)/3$, respectively. Hence, $U_1$ is a constant;
indeed, $U_1 = 4/3$.

\begin{lem}
\label{Ph1}
For each $t$, $u_1>u_2>u_3$, 
$$lim_{\xi \rightarrow 0} \Phi_1(\xi, u_1,u_2,u_3) = - \infty \ , ~~~~~
lim_{\xi \rightarrow 1} \Phi_1(\xi, u_1,u_2,u_3) = + \infty \ .$$
\end{lem} 
Proof: Since $U_1$ is a constant, we see from (\ref{split3}) that it suffices to prove that 
\begin{equation}
\label{F1L}
lim_{\xi \rightarrow 0} F_1(\xi, u_1,u_2,u_3) = + \infty \ , ~~~~~
lim_{\xi \rightarrow 1} F_1(\xi, u_1,u_2,u_3) = - \infty \ .
\end{equation}

We first  
write $F_1(\xi, u_1, u_2, u_3)$ in terms of
$F_0(\xi, u_1)$. We see from
(\ref{ph1}) that
$F_1$ satisfies
\begin{displaymath}
2(u_2 - u_3) {\p^2 F_1 \over \p u_2 \p u_3} = {\p F_1 \over \p u_2}
- {\p F_1 \over \p u_3}
\end{displaymath}
and that it has the boundary values $F_1(\xi, u_1, u, u) = {F_0(\xi, u_1)
- F_0(u, u_1) \over \xi - u}$ on account of (\ref{lim1}).
It then follows from formula (\ref{EPQsS}) at $\rho=1$ that
\begin{equation*}
F_1(\xi, u_1,u_2,u_3) = {1 \over \pi} \int_{u_3}^{u_2}
{F_0(\xi, u_1) - F_0(u, u_1) \over (\xi -u) \sqrt{(u_2 -u)(u - u_3)}}
d u \ ,
\end{equation*}
which together with (\ref{F}) gives (\ref{F1L}). This proves Lemma \ref{Ph1}.

\begin{lem}
\label{m=2}
Under the conditions (\ref{cf1}) and (\ref{cf2}), 
$\Phi_1(\xi, u_1,u_2,u_3)$ has at most one $\xi$-zero for all $t$, $u_1$, $u_2$ and
$u_3$ whenever $1 - \delta < u_1 < 1$, where $\delta$ is given by Theorem \ref{leading3}.
\end{lem}
Proof:  We first observe that
\begin{equation}
\label{unique}
{\p \over \p \xi} F_1(\xi, u_1,u_2,u_3) < 0
\end{equation}
for all $\xi$, $u_1$, $u_2$ and $u_3$ whenever $1 - \delta < u_1 < 1$.
To see this, we will derive a new formula for $F_1$ in terms of $F_0$. For each fixed $u_1$, 
we view $F_1(\xi, u_1, u_2, u_3)$
as a function of $\xi$, $u_2$ and $u_3$ only. Hence, equations (\ref{ph1}) and (\ref{ph2}) for $F_1$
reduce to
\begin{equation}
\label{new1}
2(u_2 - u_3) {\p^2 F_1 \over \p u_2 \p u_3} = {\p F_1 \over \p u_2} - {\p F_1 \over \p u_3} \ ,
\ 2(\xi - u_i) {\p^2 F_1 \over \p \xi \p u_i} = {\p F_1 \over \p \xi} - 2 {\p F_1 \over \p u_i} \ , \ i = 2 ,3 \ .
\end{equation}
The new boundary condition is
\begin{equation}
\label{new2}
F_1(u, u_1, u, u) = {\p F_0(\xi, u_1) \over \p \xi}|_{\xi = u}
\end{equation}
on account of (\ref{lim2}).  
One can then use the method of Section 2 to derive a double integral formula, similar to (\ref{phi}), 
for the solution of 
equations (\ref{new1}-\ref{new2}). Indeed, we have
\begin{eqnarray*}
\lefteqn{F_1(\xi, u_1,u_2,u_3)} \no  \\
&=& {1 \over 2 \sqrt{2} \pi} \int_{-1}^1 \int_{-1}^1
{F_{0 \xi}({1 + \mu \over 2}{1 + \nu \over 2} \xi + {1 + \mu \over 2}{1 - \nu \over 2} u_2
+{1 - \mu \over 2} u_3, u_1) \over \sqrt{(1 - \mu)(1 - \nu)}} \sqrt{1 + \mu} \ d \mu d \nu \ , 
\end{eqnarray*}
which, when combined with (\ref{F_0''}), gives (\ref{unique}).

Since $U_1$ is a constant, we deduce from
the split (\ref{split3}) that
\begin{equation}
\label{u2}
{\p \over \p \xi} \Phi_1(\xi, u_1,u_2,u_3) = - {\p \over \p \xi} F_1(\xi, u_1,u_2,u_3) > 0 
\quad \mbox{for $1 - \delta < u_1 < 1$} \ ,
\end{equation}
where the inequality follows from (\ref{unique}). 
This proves the
uniqueness of the $\xi$-zero of $\Phi_1(\xi, u_1,u_2,u_3)$ and hence Lemma \ref{m=2}.

\noindent
{\bf Remark:} \ {\em This is the third and also the last place where $m=2$ is important.
If $m > 2$, (\ref{u2}) is not true.} 

We now continue to extend the solution of equations (\ref{P2'}) by decreasing $u_2$.

By Lemma \ref{Ph1}, the function $\Phi_1(\xi, u_1,u_2,u_3)$ must have at least a $\xi$-zero.
Indeed, it 
has a zero, denoted by $u^*$, between $u_3$ and $u_2$ in view of equation
(\ref{l2}). According to Lemma \ref{m=2}, $u^*$ is the only zero of $\Phi_1$.
Hence,
\begin{equation}
\label{Ph1><}
\Phi_1(\xi, u_1, u_2, u_3) > 0 \quad \mbox{for $\xi > u^*$} \ and 
\ \Phi_1(\xi, u_1, u_2, u_3) <  0 \quad \mbox{for $\xi <  u^*$} \ .    
\end{equation}
Since $u_3 <  u_2 < u_1$ and $u_3 < u^* < u_2$, we have
\begin{equation}
\label{2}
\Phi_1(u_1, u_1,u_2,u_3) > 0 \ , ~  \Phi_1(u_2, u_1,u_2,u_3) > 0 \ , ~  \Phi_1(u_3, u_1,u_2,u_3) < 0 \ .
\end{equation}

For $P_{1,0}$ of (\ref{pgn}-\ref{lc}), we also have
\begin{equation}
\label{3}
P_{1,0}(u_1, u_1,u_2,u_3) > 0 \ , ~  P_{1,0}(u_2, u_1,u_2,u_3) > 0 \ , ~  P_{1,0}(u_3, u_1,u_2,u_3) < 0 \ .
\end{equation}
To see this, we observe from (\ref{pgn}) that $P_{1,0}$ is a linear function of $\xi$ and that
it thus has only one $\xi$-zero. This zero is between $u_3$ and $u_2$ on account of equation
(\ref{lc}). Inequalities (\ref{3}) are then justified.

By Lemma \ref{jaco}, we deduce from inequalities (\ref{2}) that the Jacobian of (\ref{P2'}) is always non-zero. 
It then follows from the
decomposition (\ref{1}) for $P$ and inequalities (\ref{3}) that equations (\ref{P2'})
can be inverted to give $x$, $u_1$, and $u_3$ as increasing, decreasing, and decreasing functions of $u_2$,
respectively, provided that $u_3 <  u_2 < u_1$ and $1 - \delta < u_1 < 1$. 
We now extend the solution as far as possible by decreasing
$u_2$; so $u_1$ always stays in the interval $(1 - \delta, 1)$ because $u_1$ increases as $u_2$ decreases.
Since both $u_1$ and $u_3$ increase as $u_2$ decreases, the solution
will stop at $u_2$ = $u_3$. At this point, $x$, $u_1$, $u_2 = u_3$ satisfy the trailing edge equations
(\ref{teg10}-\ref{teg30}). According to Lemma \ref{trailing}, the equations have a unique
solution for $t > T$. We thus have $x=x^-(t)$, $u_1= u_1^-(t)$, $u_2=u_3=u_3^-(t)$.

We have therefore proved that equations (\ref{P2'}) can be inverted to give $u_1$, $u_2$ and
$u_3$ as functions of $x$ for $x^-(t) < x < x^+(t)$. 

Inequalities (\ref{constraint}-\ref{constraint2}) are immediate consequences of (\ref{l2}) and (\ref{Ph1><}).
We have therefore established the following theorem.
\begin{theo}
\label{Whitham}
Under the conditions (\ref{cf1}) and (\ref{cf2}), for each $t > max \{ T^-, T^+ \}$, the
minimizer of the minimization problem (\ref{mini}) is supported on two disjoint intervals
when $x^-(t) < x < x^+(t)$.
\end{theo}

The main theorem of this paper, Theorem \ref{Main}, follows from Theorems \ref{Burgers2} and 
\ref{Whitham}.

\bigskip

{\bf Acknowledgments.} 
We thank Tamara Grava for valuable discussions. 
V.P. was supported in part by NSF Grant DMS-0135308. F.-R. T. was supported in part by
NSF Grant DMS-0404931 and by a John Simon Guggenheim Fellowship.

\end{document}